\documentclass[usenatbib]{mn2e}
\usepackage{epsfig}
\usepackage{amssymb}
\usepackage{lscape,graphicx}
\usepackage{longtable}
\usepackage{psfig}

\usepackage{lscape}

\usepackage{rotating}

\usepackage{Times}
\usepackage{graphicx}
\usepackage{epstopdf}

\def\gsim{\mathrel{\raise0.35ex\hbox{$\scriptstyle >$}\kern-0.6em
\lower0.40ex\hbox{{$\scriptstyle \sim$}}}}
\def\lsim{\mathrel{\raise0.35ex\hbox{$\scriptstyle <$}\kern-0.6em
\lower0.40ex\hbox{{$\scriptstyle \sim$}}}}

\begin{document}
\vspace{-10cm}
\title[CO(2--1) in an SMG at $z$\,=\,4.76]
{Detection of molecular gas in a distant submillimetre galaxy at $z=4.76$ with ATCA}
\vspace{-1cm}
\author[Coppin et al.]{ 
\parbox[t]{\textwidth}{
K.E.K. Coppin,$^{\! 1}$ S.C. Chapman,$^{\! 2}$ Ian Smail,$^{\! 1}$ A.M. Swinbank,$^{\! 1}$ F. Walter,$^{\! 3}$ J.L. Wardlow,$^{\! 4}$ A. Weiss,$^{\! 5}$ D.M Alexander,$^{\! 4}$ W.N. Brandt,$^{\! 6}$ H. Dannerbauer,$^{\! 7}$ C. De Breuck,$^{\!8}$ M. Dickinson$^{\! 9}$ J.S. Dunlop$^{\! 10}$ A.C. Edge,$^{\! 1}$  B.H.C. Emonts,$^{\! 11}$ T.R. Greve,$^{\! 12}$ M. Huynh,$^{\! 13}$ R.J. Ivison,$^{\! 10,14}$ K.K. Knudsen,$^{\! 15}$ K.M. Menten,$^{\! 5}$ E. Schinnerer,$^{\! 3}$ P.P. van der Werf$^{\! 16}$}\\\\
$^{1}$ Institute for Computational Cosmology, Durham University, South Road, Durham, DH1 3LE, UK\\
$^{2}$ Institute of Astronomy, Madingley Road, Cambridge, CB3 0HA, UK \\
$^{3}$ Max-Planck-Institut f\"{u}r Astronomie, K\"{o}nigstuhl 17, Heidelberg, D-69117, Germany\\
$^{4}$ Department of Physics, Durham University, South Road, Durham, DH1 3LE, UK \\
$^{5}$ Max-Planck-Institut f\"{u}r Radioastronomie, Auf dem H\"{u}gel 69, Bonn, D-53121, Germany \\
$^{6}$ Department of Astronomy and Astrophysics, Pennsylvania State University, 525 Davey Lab, University Park, PA 16802, USA\\
$^{7}$ Laboratoire AIM, CEA/DSM-CNRS-Université Paris Diderot, DAPNIA/Service d'Astrophysique, CEA Saclay, Orme de Merisiers, 91191 Gif-sur-Yvette, Cedex, France\\
$^{8}$ European Southern Observatory, Karl-Schwarzschild Strasse, 85748 Garching bei M\"{u}nchen, Germany\\
$^{9}$ National Optical Astronomy Observatory, P.O.\ Box 26732, Tucson, AZ 85726, USA \\
$^{10}$ SUPA, Institute for Astronomy, University of Edinburgh, Royal Observatory, Blackford Hill, Edinburgh, EH9 3HJ, UK\\
$^{11}$ Australia Telescope National Facility, CSIRO Astronomy and Space Science, PO Box 76, Epping, NSW 1710, Australia\\
$^{12}$ Dark Cosmology Centre, Niels Bohr Institute, University of Copenhagen, Juliane Maries Vej 30, DK-2100 Copenhagen, Denmark\\
$^{13}$ Infrared Processing and Analysis Center, MS220-6, California Institute of Technology, Pasadena, CA 91125, USA\\
$^{14}$ UK Astronomy Technology Centre, Royal Observatory, Blackford Hill, Edinburgh, EH9 3HJ, UK \\
$^{15}$ Argelander Institute for Astronomy, University of Bonn, Auf dem H\"{u}gel 71, D-53121 Bonn, Germany \\
$^{16}$ Leiden Observatory, Leiden University, PO Box 9513, NL-2300 RA Leiden, the Netherlands \\
}
\maketitle
\vspace{-5cm}
\begin{abstract}
We have detected the CO(2--1) transition from the submillimetre galaxy (SMG) LESS\,J033229.4$-$275619 at $z=4.755$ using the new Compact Array Broadband Backend system on the Australian Telescope Compact Array.  These data have identified a massive gas reservoir available for star formation for the first time in an SMG at $z\sim5$.  We use the luminosity and velocity width (FWHM of $\simeq160$\,km\,s$^{-1}$) of the CO(2--1) line emission to constrain the gas and dynamical mass of M$_\mathrm{gas}\simeq 1.6\times10^{10}$\,M$_\odot$ and  M$_\mathrm{dyn}(<2\,\mathrm{kpc})\simeq 5\times 10^{10}$\,(0.25/sin$^{2}i$)\,M$_{\odot}$, respectively, similar to that observed for SMGs at lower redshifts of $z\sim2$--4, although we note that our observed CO FWHM is a factor of $\sim3$ narrower than typically seen in SMGs.  Together with the stellar mass we estimate a total baryonic mass of M$_\mathrm{bary}\simeq 1\times10^{11}$\,M$_\odot$, consistent with the dynamical mass for this young galaxy within the uncertainties.  Dynamical and baryonic mass limits of high-redshift galaxies are useful tests of galaxy formation models:  using the known $z\sim4$--5 SMGs as examples of massive baryonic systems, we find that their space density is consistent with that predicted by current galaxy formation models.  In addition, these observations have helped to confirm that $z\sim4$--5 SMGs possess the baryonic masses and gas consumption timescales necessary to be the progenitors of the luminous old red galaxies seen at $z\sim3$.  Our results provide a preview of the science that ALMA will enable on the formation and evolution of the earliest massive galaxies in the Universe.
\end{abstract}


%
%
%
\section{Introduction}
Blank field surveys at millimeter and submillimetre wavelengths have uncovered a population of extremely luminous, highly optically obscured star-forming galaxies, so termed submillimetre galaxies (SMGs; e.g.~\citealt{Smail97}; \citealt{Hughes98}; \citealt{Barger98}; \citealt{Coppin06}; \citealt{Weiss09}).  SMGs are infrared luminous (L$_\mathrm{IR}\gsim5\times10^{12}$\,L$_\odot$) and have high inferred star formation rates (SFRs) in excess of 1000\,M$_{\odot}$\,yr$^{-1}$, and the redshift distribution of radio-identified 850-$\mu$m selected SMGs (representing $\approx 60$\% of the SMG population; e.g.~\citealt{Ivison07}) peaks at $z\sim 2.2$ \citep{Chapman05}.  There are indications that SMGs are strongly clustered (\citealt{Webb03}; \citealt{Blain04}; \citealt{Weiss09}), while molecular gas observations of SMGs have demonstrated that they contain the massive compact gas reservoirs needed to form a luminous spheroid (\citealt{Greve05}; \citealt{Tacconi06,Tacconi08,Tacconi10}).  The combination of these properties make them attractive candidates to be progenitors of today's most massive elliptical galaxies (e.g.~\citealt{Lilly99}; \citealt{Genzel03}; \citealt{Swinbank06}). 

While the bulk of SMGs lie at $z\sim2$, recently the first examples of submm-selected SMGs at $z\sim 4$--5 have been uncovered (\citealt{Capak08}; \citealt{Schinnerer08}; \citealt{Daddi09a,Daddi09b}; \citealt{Coppin09}; \citealt{Knudsen10}), reviving the potential worry that significant numbers of SMGs could lie at $z>4$ and are missed from current spectroscopic surveys (e.g.~\citealt{Chapman05}; \citealt{Pope05}).  The presence of even a small number of galaxies at $z\geq 4$ forming stars at a rate of $\sim 10^3$\,M$_\odot$\,yr$^{-1}$ could pose a serious problem for current hierarchical models, which predict a $z>4$ SMG surface density which is barely consistent with highly incomplete current observational estimates ($\sim10$\,deg$^{-2}$; see \citealt{Coppin09}). However, the galaxy luminosities are not robust predictions of the models since they depend upon the adopted initial mass function (IMF) and dust properties (e.g.~\citealt{Swinbank08}).  Much stronger and more reliable constraints on the models are provided by comparisons of the baryonic (gas+stars) and dynamical masses for these young galaxies (e.g.~\citealt{Genzel03}; \citealt{Baugh05}). 

In this letter we report on the detection of CO(2--1) emission from a luminous SMG, LESS\,J033229.4--275619 (LESS\,J033229.4 hereafter), discovered in the deep 870-$\mu$m Large Apex BOlometer CAmera (LABOCA) Extended Chandra Deep Field South (ECDFS) Survey (LESS; \citealt{Weiss09}).  LESS\,J033229.4 has a deboosted 870-$\mu$m flux density of $5.1\pm1.4$\,mJy (4.6\,$\sigma$ significance), a faint radio and mid-infrared counterpart, and optical spectroscopic and photometric observations identifying it as an ultraluminous far-infrared galaxy at $z=4.76$ (see \citealt{Coppin09}).   This is currently the highest redshift SMG known, suggesting that intense star formation was occurring in galaxies when the Universe was just over 1\,Gyr old (see also \citealt{Riechers09}).   Here, we use observations of CO(2--1) to a) confirm the counterpart identification and redshift of the source; b) to investigate the L$'_\mathrm{CO}$--L$_\mathrm{IR}$ relation for SMGs at $z\sim5$; c) to determine if $z>4$ SMGs have the baryonic and gas consumption time scales necessary to be the progenitors of the luminous red galaxies found at $z\sim3$ (\citealt{Marchesini07,Marchesini09}); and d) to derive a total baryonic mass for the system by combining our stellar and gas mass estimates, in order to test the \citet{Baugh05} galaxy formation model.   We adopt cosmological parameters of $\Omega_\Lambda=0.73$, $\Omega_\mathrm{m}=0.27$, and $H_\mathrm{0}=71$\,km\,s$^{-1}$\,Mpc$^{-1}$.  At $z=4.76$ the angular scale is 6.5\,kpc arcsec$^{-1}$, and the Universe is 1.3\,Gyr old.

\vspace{-0.5cm}
\section{Observations and data reduction}
The $^{12}$CO(2--1) line ($\nu_\mathrm{rest}=230.538$\,GHz) tracing cold molecular gas in LESS\,J033229.4 ($\alpha_{\rm J2000}= 03^\mathrm{h}32^\mathrm {m}29.30^\mathrm{s}$, $\delta_{\rm J2000}=-27^{\circ}56^{'}19.40^{''}$) was observed in 2009 September over a period of 6 nights with the Australia Telescope Compact Array (ATCA), fitted out with the recently upgraded Compact Array Broadband Backend (CABB; \citealt{Ferris02}).  For improved ($u,v$)-coverage, we employed a combination of the two most sensitive and compact 5-antenna hybrid configurations, H75 and H168.
We utilised the maximum available CABB 2.048\,GHz bandwidth (ensuring that the CO line of interest would lie well within the spectral window), with 2048 channels across the intermediate-frequency (IF) band, corresponding to a 1\,MHz spectral resolution.   Assuming a source redshift of $z=4.762$ \citep{Coppin09}, we tuned the 7\,mm IF1 receiver to a central frequency of 40.010\,GHz, resulting in a frequency coverage of 39.035--40.984\,GHz, covering CO(2--1) emission between $z=4.625$--4.906.  The system temperature ranged from 90--160\,K over the observing run.  Phase, amplitude, and bandpass calibration were acquired by short observations of PKS\,0402--362 every 10 mins, with pointing performed on the same source every hour.  Flux density calibration was performed using short observations of Uranus each night (and Mars in one instance), the recommended standard ATCA calibrator source, and we estimate the calibration uncertainty to be $\lsim 20$\% based on the currently available standard MIRIAD model for Uranus (\citealt{Sault99}; see \citealt{Emonts10}).

The data were calibrated, mapped and analysed using the standard \textsc{Miriad}\footnote{http://www.atnf.csiro.au/computing/software/miriad} \citep{Sault99} and \textsc{Karma}\footnote{http://www.atnf.csiro.au/computing/software/karma} \citep{Gooch96} packages.  The corresponding synthesized beam 
adopting a natural weighting is $\approx12 \times 9$\,arcsec$^{2}$.
Overall 35\,hr of observations were obtained, including 25\,hr of on-source integration time.
The resultant 1\,$\sigma$ rms channel noise in the 1\,MHz-wide channel spectrum (corresponding to $\simeq7$\,km\,s$^{-1}$) 
is $\approx0.44$\,mJy\,beam$^{-1}$, which is consistent with sensitivities reached by other 7\,mm ATCA/CABB experiments (e.g.~\citealt{Emonts10}).

\vspace{-0.5cm}
\section{Observed Molecular Gas Properties}
\begin{figure}
\epsfig{file=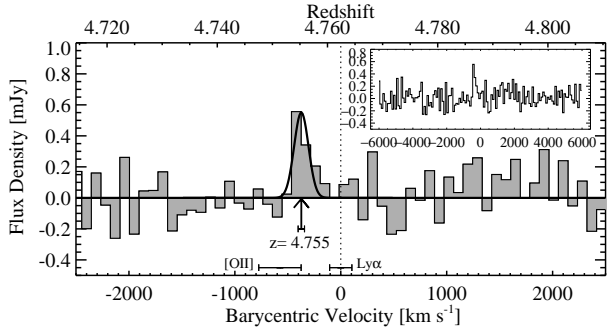,width=0.5\textwidth}
\vspace{-5cm}
\caption{CO(2--1) spectrum of LESS\,J033229.4 binned into 90\,km\,s$^{-1}$ channels, extracted through the brightest pixel from Fig.~\ref{fig:chan}, showing a $\simeq5\,\sigma$ line detection overlaid with a best-fitting Gaussian with a FWHM of $160\pm65$\,km\,s$^{-1}$.  The upper axis indicates the redshift of the CO(2--1) transition, from which the line emission fit indicates a $z=4.755\pm0.001$.  The CO emission line is blueshifted by $\approx 370$\,km\,s$^{-1}$ from the optical Ly$\alpha$ redshift (dotted vertical line and error bar) and consistent with a redshift of $z=4.751\pm0.005$ derived from a VLT [OII] emission spectrum (Alaghband-Zadeh et al.\ in preparation).  We have zoomed in on the central $\sim5000$\,km\,s$^{-1}$ of the spectrum for clarity, although for completeness we also show the entire 2\,GHz wide spectrum as an inset.}
\label{fig:spec}
\end{figure}

\begin{figure}
\epsfig{file=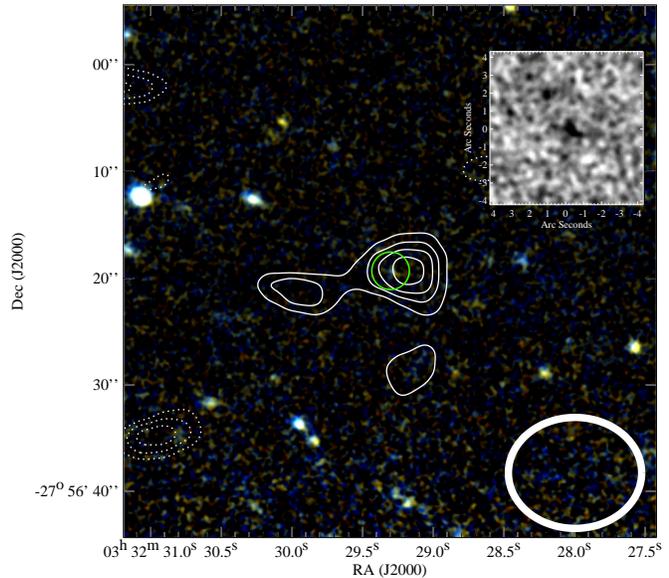,width=0.5\textwidth}
\caption{CO(2--1) intensity contours of the line peak in a 90\,km\,s$^{-1}$-wide channel ATCA spectrum overlaid on a true-colour ground-based HAWK-I JK image of LESS\,J033229.4.  The CO(2--1) solid line contours begin at 2.5\,$\sigma$ and increase in steps of 0.5\,$\sigma$, where $\sigma=0.14$\,mJy (the channel rms). Equivalent negative contours are shown by dotted curves.  The CO source appears to be spatially unresolved with the $\approx12\times9$\,arcsec$^{2}$ beam ($\simeq 80\times60$\,kpc FWHM; shown in the bottom right-hand corner). The CO centroid is consistent with the radio-identified optical $V$-band dropout counterpart of the SMG (encircled in green and featured as an inset in the top right-hand corner) within the $\lsim 2.3$\,arcsec positional uncertainty.}
\label{fig:chan}
\end{figure}

Inspection of the combined data cube reveals the presence of a relatively narrow emission line near the phase centre, blueshifted by $\sim400$\,km\,s$^{-1}$ from the central frequency.  To test the reality of the detection, we split the visibility data into two halves and find that the line persists in the data cubes at lower signal-to-noise.  To quantify the significance and parameters of the CO(2--1) line emission, the visibilities were resampled to a velocity resolution of 90\,km\,s$^{-1}$ (chosen to yield optimal line SNR), providing a 1\,$\sigma$ rms channel sensitivity of $\simeq0.14$\,mJy\,beam$^{-1}$, revealing a $\simeq5\,\sigma$ detection (see Fig.~\ref{fig:spec}),  the brightest peak in the entire data cube with this binning.  In Fig.~\ref{fig:chan}, we plot contours of the emission for the line at the brightest 90\,km\,s${-1}$ channel overlaid on a true-colour near-infrared HAWK-I $JK$ image of LESS\,J033229.4 from Zibetti et al.\ (in preparation).  A Gaussian profile fit to the CO spectrum indicates a full width at half maximum (FWHM) of $160\pm65$\,km\,s$^{-1}$, and a velocity integrated line flux of F$_\mathrm{CO(2-1)}=0.09\pm0.02$\,Jy\,km\,s$^{-1}$.  The uncertainties on the CO line parameters have been estimated by rebinning the data cube to several different spectral channel resolutions from 50--200\,km\,s$^{-1}$ and finding the 1\,$\sigma$ spread of each parameter, which are comparable to the individual parameter fit uncertainties.  We determine a best-fitting CO(2--1) redshift of $z=4.755\pm0.001$, corresponding to an offset of $-370\pm30$km\,s$^{-1}$ with respect to the optical redshift derived by \citet{Vanzella06} of $z=4.762\pm0.002$ determined from the asymmetric Ly$\alpha$ emission, N{\sc v} and a continuum break (see \citealt{Coppin09} for the optical spectrum).  Note that this apparent redshifting of Ly$\alpha$ compared to the systemic redshift of the source (as measured from the nebular emission) is routinely seen in samples of $2 \lsim z \lsim3$ star-forming galaxies with offsets of $<\! \Delta \mathrm{v}_{\mathrm{Ly}\alpha}\!>\simeq 450$\,km\,s$^{-1}$ compared to the systemic redshift \citep{Steidel10}.  The CO redshift also corresponds to the redshift of the [OII] emission at $z=4.751\pm0.005$ from a VLT spectrum (Alaghband-Zadeh et al.\ in preparation).  This CO detection has provided definitive proof that the previously identified counterpart at $z=4.76$ for this SMG by \citet{Coppin09} is correct.

Averaging the remaining channels in the $\sim1.6$\,GHz (129-channel) line-free region of the spectrum reveals no significant continuum detection down to a 1\,$\sigma$ sensitivity of 0.013\,mJy. 

\vspace{-0.5cm}
\section{Inferred Gas, Dynamical, and Total Baryonic Mass of the System}
We calculate the line luminosity and the total cold gas mass (H$_{2}$+He) from the integrated CO(2--1) line flux following \citet{Solomon05} in order to determine the amount of available fuel for the current star formation episode.  We find a line luminosity of L$'_\mathrm{CO(2-1)}=(2.0\pm0.4)\times10^{10}$\,K\,km\,s$^{-1}$\,pc$^{2}$.  We then assume both that the gas is thermalised for the $J=2$--1 transition (i.e.\ assuming a constant line temperature brightness ratio, L$'_\mathrm{CO}(2$--1)=L$'_\mathrm{CO}(1$--0)) and a CO-to-H$_{2}$ conversion factor of $\alpha=0.8$\,M$_{\odot}$(K\,km\,s$^{-1}$\,pc$^{2}$)$^{-1}$ (which is appropriate for ultraluminous infrared galaxies and SMGs; e.g.~\citealt{Downes98}; \citealt{Tacconi08}), yielding a total cold gas mass of M$_\mathrm{gas}$=$(1.6\pm0.3)\times10^{10}$\,M$_\odot$.  
This CO(2--1) gas mass is comparable to that derived for SMGs from higher-$J$ transitions (typically $J=3$--2 or $J=4$--3).  If LESS\,J033229.4 is a typical SMG, these CO $J=2$--1 observations suggest that we are unlikely to be missing massive reservoirs of molecular gas around SMGs (see also e.g.~\citealt{Carilli10}; \citealt{Ivison10}; cf.~\citealt{Hainline06}; \citealt{Papa10}).  

We calculate a dynamical mass of the system based on the observed CO(2--1) line width following \citet{Solomon05}:  M$_\mathrm{dyn}$\,sin$^{2}i$=233.5$\Delta\,v^{2}_\mathrm{FWHM}\times$R (M$_\odot$), where $i$ is the inclination of the gas disk, $\Delta\,v_\mathrm{FWHM}$ is the CO FWHM (in km\,s$^{-1}$), and R is the radius of the CO emitting region (in pc).  Since we do not have a direct constraint on the spatial extent of the CO emission from our unresolved detection, we assume that the CO traces the stellar light (as seems an appropriate assumption for $z\sim2$ SMGs; \citealt{Swinbank10}), which has a seeing-corrected FWHM of $0.5''\pm0.2''$ in the HAWK-I $K$-band imaging (or equivalently a FWHM $\simeq4$\,kpc at $z=4.76$).   The best-fitting Gaussian FWHM of $(160\pm40)$\,km\,s$^{-1}$ implies a strict lower limit on the dynamical mass of M$_\mathrm{dyn}(<2\,\mathrm{kpc})$\,sin$^{2}i$=($1.2\pm0.6)\times10^{10}$\,M$_\odot$, assuming a 2\,kpc radius.  Since we do not have any constraints on $i$, in this letter we assume $i=30^\circ$ (appropriate for randomly inclined disks in a sample of galaxies) and explicitly include the $i$ and R dependency in our derived value:  M$_\mathrm{dyn}(<2\,\mathrm{kpc})=(4.8\pm2.4)\times10^{10}\,(0.25/\mathrm{sin}^{2}i)(R/2\,\mathrm{kpc})$\,M$_\odot$.  We caution that M$_\mathrm{dyn}$ carries significant uncertainties which will require higher-resolution observations to accurately determine these quantities.  For instance, we note that the CO line FWHM is about a factor of 3 narrower than typically observed for $z\sim2$ SMGs \citep{Greve05}, although there are examples of SMGs and high-redshift Lyman-break galaxies (LBGs) with CO emission as narrow as our observed FWHM (e.g.~\citealt{Frayer99}; \citealt{Baker04}; \citealt{Coppin07}; \citealt{Stanway08}).  QSOs typically have narrower linewidths than SMGs by a factor of $\sim2$--3, consistent with the optically-selected QSOs being more closely inclined to the sky plane (showing its central AGN more clearly) than typical SMGs (\citealt{Greve05}; \citealt{Coppin08}; \citealt{Carilli06}).  Thus the CO line narrowness of LESS\,J033229.4 (whose {\it Hubble Space Telescope (HST)} rest-frame UV compact morphology and strong N{\sc v} emission indicates that it hosts an AGN; \citealt{Coppin09}) compared with other SMGs could indicate that $i\simeq 10^\circ$.  Assuming that LESS\,J033229.4 is more face-on, adopting $i\simeq 10^\circ$, yields M$_\mathrm{dyn}(<2\,\mathrm{kpc})=(4.0\pm2.0)\times10^{11}$\,M$_\odot$.   However, the line width differences could also indicate that LESS\,J033229.4 is at a later stage of a merger or has a different galaxy mass or size than typical $z\sim2$ SMGs, which could be explored with high-resolution imaging of the gas distribution (e.g.~\citealt{Greve05}; \citealt{Carilli06}).  If instead we assume that the gas configuration is spherical (with a uniform distibution) rather than disk-like, then the implied enclosed dynamical mass would be M$_\mathrm{dyn}(<2\,\mathrm{kpc})=(1.1\pm0.6)\times10^{10}$\,M$_\odot$.

The combination of the gas and stellar mass estimates for LESS\,J033229.4 can be used to calculate the total baryonic mass of the system.  The stellar mass of the system from \citet{Coppin09} has been reconfirmed by \citet{Wardlow10} to be $\lsim 1 \times 10^{11}$\,M$_\odot$.  \citet{Wardlow10} has included the new HAWK-I $J$ and $K$-band photometry, and adjusted the derived mass for a \citet{Salpeter55} IMF for compatibility with the IMF assumed in deriving the SFR in \S~\ref{compare}.
We caution that this quantity has an uncertainty of a factor of $\lesssim 5$ even before considering a potential AGN contribution to the rest-frame near-infrared emission and so should be considered an approximate upper limit.  Taken together, these estimates imply a total baryonic mass within 2\,kpc of M$_\mathrm{bary}$=M$_\mathrm{gas}$+M$_\mathrm{stars}=1.2^{+4.0}_{-0.8} \times 10^{11}$\,M$_{\odot}\, \simeq 1 \times 10^{11}$\,M$_\odot$, which is consistent with the dynamical mass estimated above from the CO emission given the considerable uncertainties.

\vspace{-0.5cm}
\section{Discussion and Conclusions}
\subsection{$z\sim4$--5 SMGs are higher redshift analogues of $z\sim2$ SMGs}\label{compare}
How representative is LESS\,J033229.4 of the few known $z\sim4$--5 submillimetre-selected sources and the more abundant $z\sim2$ SMG population?  Based on the photometric constraints for LESS\,J033229.4 which trace an SED consistent with local star formation dominated ULIRGs, \citet{Coppin09} derive a dust mass estimate of $M_\mathrm{d}\sim 5 \times 10^{8}$\,M$_{\odot}$ and a far-infrared luminosity of L$_\mathrm{IR}\simeq 6\times10^{12}$\,L$_\odot$, implying a SFR\footnote{The implied SFR is calculated following \citet{Kennicutt98}, assuming a 100\,Myr burst lifetime and a \citet{Salpeter55} initial mass function.} of $\simeq1000$\,M$_\odot$\,yr$^{-1}$, close to the median luminosity of the $z\sim2$ and $z>4$ SMG populations (e.g.~\citealt{Kovacs06}).  When combined with our gas mass estimate above it follows that LESS\,J033229.4 has a star formation efficiency (SFE) of $\approx 250$\,L$_\odot$\,(K\,km\,s$^{-1}$\,pc$^{2})^{-1}$, a gas-to-dynamical mass fraction of $f$=M$_\mathrm{gas}$/M$_\mathrm{dyn}\sim0.3 (\mathrm{sin}^{2}i/0.25)$, and a gas-to-dust mass ratio of $\sim30$. Assuming that the star formation follows the $K$-band light, we estimate a star formation surface density of $\Sigma_\mathrm{SFR}\sim100$\,M$_\odot$\,yr$^{-1}$\,kpc$^{-2}$, which is similar to the intense central starburst mode inferred for SMGs at $z\sim2$ (e.g.~\citealt{Smail03}).

The source properties of LESS\,J033229.4 derived from the CO line emission are very similar to those of the numerous SMGs at $z\sim2$ ($<\!$\,M$_\mathrm{gas}\!>\sim3.0\times10^{10}$\,M$_\odot$, $<\!$\,M$_\mathrm{dyn}\!>\sim1.2 \times 10^{11}$\,M$_\odot$, $f\simeq0.25$, with SFEs of $\sim450\pm170$\,L$_\odot$\,(K\,km\,s$^{-1}$\,pc$^{2})^{-1}$ and M$_\mathrm{gas}$/M$_\mathrm{dust}\sim60$; \citealt{Greve05}; \citealt{Kovacs06}) and also the rarer emerging high-redshift tail $z>4$ SMGs (\citealt{Schinnerer08}; \citealt{Daddi09a,Daddi09b}).  Combining the samples of CO-detected $z>4$ SMGs, it follows that they lie within the scatter of the L$_\mathrm{IR}$--L$'_\mathrm{CO}$ relation from \citet{Greve05} for $z\sim2$ SMGs (assuming a constant line brightness ratio), with $z\sim4$--5 SMGs spanning a range of L$_\mathrm{IR}\sim0.6$--$3\times10^{13}$\,L$_\odot$ and L$'_\mathrm{CO}\sim2$--$6\times10^{10}$\,K\,km\,s$^{-1}$\,pc$^{2}$.  It thus appears that within the $z\sim1$--5 SMG population, L$_\mathrm{IR}$/L$'_\mathrm{CO}$ is constant with redshift, indicating that SMGs at $z\sim4$--5 are consistent with being higher-redshift analogues of SMGs at $z\sim2$, forming stars with similar efficiencies.   

Overall this comparison suggests that SMGs at $z\sim2$ and $z>4$ are equally evolved, have similar reservoirs of 
gas, similar star formation efficiencies, and similar fractions of baryons in cold gas as stars.

\vspace{-0.5cm}
\subsection{The potential descendents of $4<z<5$ SMGs}
LESS\,J033229.4 appears to be a luminous massive starburst, forming stars at a rate of $\sim1000$\,M$_\odot$\,yr$^{-1}$, with a total baryonic mass of $\simeq 1\times 10^{11}$\,M$_\odot$, with properties representative of SMGs at $z\sim2$--4.  The combination of its compact morphology and high SFR is potentially consistent with the small sizes of $\lsim1$--2\,kpc claimed for a population of massive extremely dense old quiescent $z\sim2$ galaxies  (e.g.~\citealt{Daddi05}; \citealt{Toft07,Toft09}; \citealt{Zirm07}; \citealt{Buitrago08}; \citealt{Cimatti08}).  Such compact dense galaxies at $z\sim2$ have implied formation redshifts of $z\sim4$--5 and are consistent with being the descendents of a population of gas-rich, highly dissipative mergers, such as $z>4$ SMGs.  

A critical question we can now address is whether LESS\,J033229.4 (assumed to be representative of $z>4$ SMGs) has the baryonic mass and gas consumption timescale necessary to be a `prototypical' progenitor of the luminous `red and dead' galaxy population found at $z\sim3$?  The baryonic content of LESS\,J033229.4 is roughly equivalent to the typical stellar mass of a giant elliptical galaxy ($1\times 10^{11}$\,M$_\odot$; \citealt{Marchesini09}), given the large uncertainties in some of our mass estimates.  Assuming that the molecular gas reservoir is fueling the star formation within LESS\,J033229.4, then it will have enough gas to sustain the current star formation episode for $\tau_\mathrm{depletion}\sim$M$_\mathrm{gas}$/SFR$\sim 1.6 \times 10^{10}$\,M$_\odot$/1000\,M$_\odot$ \,yr$^{-1}\sim 16$\,Myr, assuming 100\% efficiency.  We also compare $\tau_\mathrm{depletion}$ with the time to form the current stellar mass of the system.  At the current SFR, we estimate a $\tau_\mathrm{formation}\sim$M$_\mathrm{stars}$/SFR$\sim 1\times 10^{11}$\,M$_\odot$/1000\,M$_\odot$\,yr$^{-1} \sim 100$\,Myr, which is comparable to the assumed age of the stellar population and burst in the model used in \citet{Coppin09}.  Although the gas consumption timescale appears to be relatively short, it is similar to the gas depletion timescales of $\sim100$\,Myr for SMGs (e.g.~\citealt{Tacconi08}).  It follows that we are catching this SMG approximately half way through its current star formation episode, with the majority of the galaxy mass already in the form of stars, representative of a major stage in galaxy formation.  It thus seems plausible that LESS\,J033229.4 would display the observed properties of being `red and dead' by $z\sim3$ since the current stellar population would age another $>1$\,Gyr between $z=4.76$ and $z=3$, assuming that no further star formation occurs.

\vspace{-0.5cm}
\subsection{Observed mass constraints of $4<z<5$ SMGs on galaxy formation models}
\begin{figure}
\epsfig{file=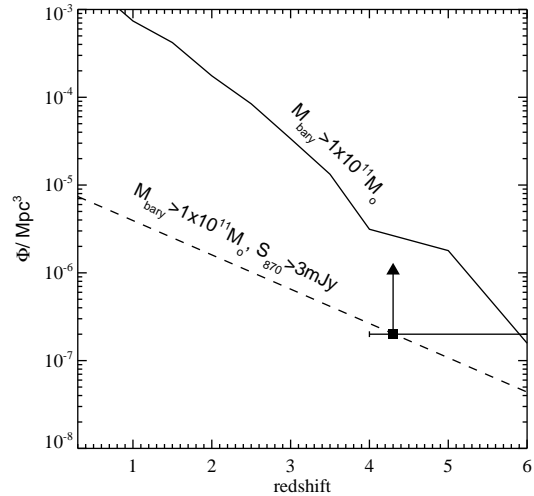,width=0.4\textwidth,angle=90}
\caption{A comparison of the theoretical predictions and observational constraints of the cosmic comoving volume density of M$_\mathrm{bary}\gsim 1\times 10^{11}$\,M$_\odot$ galaxies, an approximate baryonic mass lower limit observed for the five CO-detected $z>4$ SMGs discovered to date (the square is positioned at the mean redshift of the sample, and the horizontal line indicates the assumed redshift range $4<z<8$; see text).  The solid curve represents galaxies from the \citet{Baugh05} GALFORM semianalytic model with M$_\mathrm{bary}>1\times 10^{11}$\,M$_\odot$.  The dashed curve represents the same mass cut from the model with an additional submm flux cut applied of S$_\mathrm{850\,\mu m}>3$\,mJy (appropriate for our SMG sample), although this curve is difficult to compare directly with the observational constraints since the extraction of the submm fluxes from the model depends on the precise assumed IMF and dust properties of the model SMGs.  Thus, the solid curve demonstrates that the detailed theoretical models are consistent with current observational constraints.}
\label{fig:bary}
\end{figure}

The presence of even the small observed (and likely still incomplete) volume density of the five $z\sim4$--5 SMGs, 
$2\times10^{-7}$\,Mpc$^{-3}$ (conservatively assuming a selection function that is uniformly sensitive across $z=4$--8 due to the negative $K$-correction; e.g.~\citealt{Blain02}), poses a challenge to current hierarchical models (see \citealt{Coppin09}).  However, the prediction of submm fluxes from the models depends on the assumed IMF and dust properties in the galaxies, making a straightforward comparison between the properties of the observed and model SMGs fundamentally problematic.  A much more straightforward and reliable test is thus provided by comparisons of the total baryonic (cold gas+stars) or dynamical masses for these young galaxies (e.g.~\citealt{Genzel03}), which we now investigate using observational M$_\mathrm{bary}$ and M$_\mathrm{dyn}$ constraints of the five CO-detected $z>4$ SMGs, with a range of M$_\mathrm{bary}\sim 1$--$3\times 10^{11}$\,M$_{\odot}$ and M$_\mathrm{dyn}\sim 0.5$--$9\times 10^{11}$\,M$_{\odot}$.  In Fig.~\ref{fig:bary} we compare the observational and theoretical cosmic volume density of galaxies with M$_\mathrm{bary} \gsim 1\times 10^{11}$\,M$_\odot$ (the lower end of the distribution among the sample), using high-redshift SMGs as examples of massive baryonic galaxies with a uniform selection function out to $z\sim8$.  This exercise demonstrates in a less-biased way that the \citet{Baugh05} semianalytic GALFORM model is currently consistent with our current observed limits on the volume densities of M$_\mathrm{bary}\gsim 1\times 10^{11}$\,M$_\odot$ SMGs at $z>4$.  

These observations provide a view of a massive gas reservoir in a galaxy forming just 1\,Gyr after the Big Bang, giving a sneak preview of the capabilities of ALMA for studying the formation and evolution of the earliest galaxies.  However, statistically significant samples of spectroscopically-confirmed SMGs are required to determine the true incidence of SMGs at $z>4$, and far-infrared and (sub)millimetre continuum surveys are first needed to identify large samples of SMGs that can become the targets of interferometric follow-up efforts such as those described here.  Deep multiwavelength near-infrared imaging and sensitive (preferably) low $J$-transition CO mapping campaigns are necessary for deriving accurate stellar and gas masses, respectively:  the key to determining the contribution of SMGs to the baryonic mass of the Universe at $z>4$.

\vspace{-0.5cm}
\section{Acknowledgments}\label{ack}

We thank an anonymous referee for suggestions which improved the paper.  KEKC acknowledges support from a Science and Technology Facilities Council (STFC) fellowship.  We thank John Helly, Carlton Baugh and Cedric Lacey for help with extracting information from the Millenium and GALFORM data bases.  IRS and JLW acknowledge support from STFC, and AMS acknowledges support from a Lockyer fellowship. JSD acknowledges the support of the Royal Society via a Wolfson Research merit Award. TRG acknowledges support from IDA. The DARK Cosmology Centre is funded by DNRF.  The ATCA is part of the Australia Telescope which is funded by the Commonwealth of Australia for operation as a National Facility managed by CSIRO.  We thank everyone who made the CABB upgrades possible and also to the ATCA staff for support during the observing run.  We thank Stefano Zibetti and Nelson Padilla for the HAWK-I imaging which was obtained during ESO programme ID 082.A-0890.

\vspace{-0.5cm}
\setlength{\bibhang}{2.0em}


\begin{thebibliography}{50}
\setlength{\itemindent}{-2.5em}
\bibitem[Baker et al.(2004)]{Baker04}Baker A.J., Tacconi L.J., Genzel R., Lehnert M.D., Lutz D., 2004, ApJ, 604, 125
\bibitem[Barger et al.(1998)]{Barger98}Barger A.J., Cowie L.L., Sanders D.B., Fulton E., Taniguchi Y., Sato Y., Kawara K., Okuda H., 1998, Nat., 394, 248
\bibitem[Baugh et al.(2005)]{Baugh05}Baugh C.M., Lacey C.G., Frenk C.S., Granato G.L., Silva L., Bressan A., Benson A.J., Cole S., 2005, MNRAS, 356, 1191
\bibitem[Blain et al.(2002)]{Blain02}Blain A.W., Smail I., Ivison R.J., Kneib J.-P., Frayer D.T., 2002, Phys. Rep., 369, 111
\bibitem[Blain et al.(2004)]{Blain04} Blain A.W., Chapman S.C., Smail I., Ivison R., 2004, ApJ, 611, 725
\bibitem[Buitrago et al.(2008)]{Buitrago08}Buitrago F., Trujillo I., Conselice C.J., Bouwens R.J., Dickinson M., Yan H., 2008, ApJ, 687, 61L
\bibitem[Capak et al.(2008)]{Capak08}Capak P. et al., 2008, ApJ, 681, L53
\bibitem[Carilli \& Wang(2006)]{Carilli06}Carilli C.L. \& Wang R., 2006, ApJ, 131, 2763
\bibitem[Carilli et al.(2010)]{Carilli10}Carilli C.L. et al., 2010, ApJ, 714, 1407
\bibitem[Chapman et al.(2005)]{Chapman05}Chapman S.C., Blain A.W., Smail I., Ivison R.J., 2005, ApJ, 622, 772
\bibitem[Cimatti et al.(2008)]{Cimatti08}Cimatti A. et al., 2008, A\&A, 482, 212
\bibitem[Coppin et al.(2006)]{Coppin06}Coppin K. et al., 2006, MNRAS, 372, 1621
\bibitem[Coppin et al.(2007)]{Coppin07}Coppin K. et al., 2007, ApJ, 665, 936
\bibitem[Coppin et al.(2008)]{Coppin08}Coppin K. et al, 2008, MNRAS, 389, 45
\bibitem[Coppin et al.(2009)]{Coppin09}Coppin K. et al., 2009, MNRAS, 395, 1905 
\bibitem[Daddi et al.(2005)]{Daddi05}Daddi E. et al., 2005, ApJ, 626, 680
\bibitem[Daddi et al.(2009a)]{Daddi09a}Daddi E. et al., 2009a, ApJ, 694, 1517
\bibitem[Daddi et al.(2009b)]{Daddi09b}Daddi E., Dannerbauer H., Krips M., Walter F., Dickinson M., Elbaz D., Morrison G.E., 2009b, ApJ, 695, 176L
\bibitem[Downes \& Solomon(1998)]{Downes98} Downes D. \& Solomon P.M., 1998, ApJ, 507, 615
\bibitem[Emonts et al.(2010)]{Emonts10}Emonts B., et al., 2010, MN, submitted
\bibitem[Ferris \& Wilson(2002)]{Ferris02}Ferris R.H. \& Wilson W.E., 2002 URSI XXVIIth General Assembly, poster 1629
\bibitem[Frayer et al.(1999)]{Frayer99}Frayer D.T. et al., 1999, ApJ, 514, L13
\bibitem[Genzel et al.(2003)]{Genzel03} Genzel R., Baker A.J., Tacconi L.J., Lutz D., Cox P., 2003, ApJ, 584, 633
\bibitem[Greve et al.(2005)]{Greve05}Greve T.R. et al., 2005, MNRAS, 359, 1165
\bibitem[Gooch et al.(1996)]{Gooch96}Gooch R., 1996, ASPC, 101, 80
\bibitem[Hainline et al.(2006)]{Hainline06}Hainline L.J., Blain A.W., Greve T.R., Chapman S.C., Smail I., Ivison R.J., 2006, ApJ, 650, 614
\bibitem[Hughes et al.(1998)]{Hughes98}Hughes D.H. et al., 1998, Nat., 394, 241 
\bibitem[Ivison et al.(2007)]{Ivison07}Ivison R.J., et al., 2007, MNRAS, 380, 199
\bibitem[Ivison et al.(2010)]{Ivison10}Ivison R.J., Smail I., Papadopoulos P. P., Wold I., Richard J., Swinbank A. M., Kneib J.-P., Owen F.N., 2010, MNRAS, 404, 198
\bibitem[Kennicutt(1998)]{Kennicutt98}Kennicutt R.C., 1998, ARA\&A, 36, 189
\bibitem[Knudsen et al.(2010)]{Knudsen10}Knudsen K.K., Kneib J.-P., Richard J., Petitpas G., Egami E., 2010, ApJ, 709, 210
\bibitem[Kov\'{a}cs et al.(2006)]{Kovacs06}Kov\'{a}cs A., Chapman S.C., Dowell C.D., Blain A.W., Ivison R.J., Smail I., Phillips T.G., 2006, ApJ, 650, 592
\bibitem[Lilly et al.(1999)]{Lilly99}Lilly S.J., Eales S.A., Gear W.K.P., Hammer F., Le F\`{e}vre O., Crampton D., Bond J.R., Dunne L., 1999, ApJ, 518, 641
\bibitem[Marchesini et al.(2007)]{Marchesini07}Marchesini D. et al., 2007, ApJ, 656, 42
\bibitem[Marchesini et al.(2009)]{Marchesini09}Marchesini D., van Dokkum P.G., F\"{o}rster-Schreiber N.M., Franx M., Labb\'{e} I., Wuyts S., 2009, ApJ, 701, 1765
\bibitem[Papadopoulos et al.(2010)]{Papa10}Papadopoulos P.P., van der Werf P., Isaak K., Xilouris E.M., 2010, ApJ, 715, 775
\bibitem[Pope et al.(2005)]{Pope05}Pope A., Borys C., Scott D., Conselice C., Dickinson M., Mobasher B., 2005, MNRAS, 358, 149
\bibitem[Salpeter(1955)]{Salpeter55}Salpeter E.E., 1955, ApJ, 121, 161
\bibitem[Sault \& Killeen(1999)]{Sault99}Sault R.J. \& Killeen N.E.B., 1999, MIRIAD User's Guide (Sydney: Australia Telescope National Facility)
\bibitem[Schinnerer et al.(2008)]{Schinnerer08}Schinnerer E. et al., 2008, ApJ, 689, 5L
\bibitem[Smail, Ivison \& Blain(1997)]{Smail97}Smail I., Ivison R.J. \& Blain A.W., 1997, ApJ, 490, 5L
\bibitem[Smail et al.(2003)]{Smail03}Smail I., Chapman S.C., Ivison R.J., Blain A.W., Takata T., Heckman T.M., Dunlop J.S., Sekiguchi K., 2003, MNRAS, 342, 1185
\bibitem[Solomon \& Vanden Bout(2005)]{Solomon05}Solomon P.M. \& Vanden Bout P.A., 2005, ARA\&A, 43, 677
\bibitem[Swinbank et al.(2006)]{Swinbank06} Swinbank A.M., Chapman S.C., Smail I., Lindner C., Borys C., Blain A.W., Ivison R.J., Lewis G.F., 2006, MNRAS, 371, 465
\bibitem[Swinbank et al.(2008)]{Swinbank08} Swinbank A.M. et al., 2008, MNRAS, 391, 420
\bibitem[Swinbank et al.(2010)]{Swinbank10}Swinbank A.M. et al., 2010, MNRAS, 405, 234
\bibitem[Stanway et al.(2008)]{Stanway08} Stanway E.R., Bremer M.N., Davies L.J.M., Birkinshaw M., Douglas L.S., Lehnert M.D., 2008, ApJ, 687, 1L
\bibitem[Steidel et al.(2010)]{Steidel10}Steidel C.C., Erb D.K., Shapley A.E., Pettini M., Reddy N.A., Bogosavljevi\'{c} M., Rudie G.C., Rakic O., 2010, ApJ, 717, 289
\bibitem[Riechers et al.(2009)]{Riechers09}Riechers D.A. et al., 2009, ApJ, 703, 1338
\bibitem[Tacconi et al.(2006)]{Tacconi06}Tacconi L.J. et al., 2006, ApJ, 640, 228
\bibitem[Tacconi et al.(2008)]{Tacconi08}Tacconi L.J. et al., 2008, ApJ, 680, 246
\bibitem[Tacconi et al.(2010)]{Tacconi10}Tacconi L.J. et al., 2010, Nat., 463, 781
\bibitem[Toft et al.(2007)]{Toft07}Toft S., et al, 2007, ApJ, 671, 285
\bibitem[Toft et al.(2009)]{Toft09}Toft S., Franx M., van Dokkum P., F\"{o}rster Schreiber N.M., Labb\'{e} I., Wuyts S., Marchesini D., 2009, ApJ, 705, 255
\bibitem[Vanzella et al.(2006)]{Vanzella06}Vanzella E. et al., 2006, A\&A, 454, 423
\bibitem[Wardlow et al.(2010)]{Wardlow10}Wardlow J.L. et al., 2010, preprint (arXiv:1006.2137)
\bibitem[Webb et al.(2003)]{Webb03}Webb T.M. et al., 2003, ApJ, 587, 41
\bibitem[Wei\ss\ et al.(2009)]{Weiss09}Wei\ss\ A. et al., 2009, MNRAS, 707, 1201
\bibitem[Zirm et al.(2007)]{Zirm07}Zirm A.W. et al., 2007, ApJ, 656, 66
\end{thebibliography}
\end{document}